%% file: Forensic_Similarity.tex
\documentclass[sigconf]{acmart}

\AtBeginDocument{%
  }

\copyrightyear{2026}
\acmYear{2026}
\setcopyright{cc}
\setcctype{by}
\acmConference[IH\&MMSec '26]{ACM Workshop on Information Hiding and Multimedia Security}{June 17--19, 2026}{Firenze, Italy}
\acmBooktitle{ACM Workshop on Information Hiding and Multimedia Security (IH\&MMSec '26), June 17--19, 2026, Firenze, Italy}
\acmDOI{10.1145/3785353.3815078}
\acmISBN{979-8-4007-2376-6/2026/06}

\acmConference[IH\&MMSec ’26]{Proceedings of the 2026 ACM Workshop on Information
Hiding and Multimedia Security}{June 17–19, 2026}{Florence, Italy}

\acmISBN{}

\usepackage{booktabs}
\usepackage{tabularx}
\usepackage{longtable} 
\usepackage{colortbl}
\usepackage{multirow, multicol, array, makecell}
\usepackage{siunitx}
\usepackage{cleveref}
\usepackage{flushend}
\crefname{section}{Section}{Sections}
\crefname{figure}{Figure}{Figures}
\crefname{table}{Table}{Tables}

\usepackage{glossaries}
\newacronym{cnn}{CNN}{Convolutional Neural Network}
\newacronym{dl}{DL}{Deep Learning}
\newacronym{ai}{AI}{Artificial Intelligence}
\newacronym{dnn}{DNN}{Deep Neural Network}
\newacronym{nn}{NN}{Neural Network}
\newacronym{mmf}{MMF}{MultiMedia Forensics}
\newacronym{vc}{VC}{Voice Conversion}
\newacronym{tts}{TTS}{Text-to-Speech}
\newacronym{ann}{ANN}{Artificial Neural Network}
\newacronym{df}{DF}{Deepfake}
\newacronym{ml}{ML}{Machine Learning}
\newacronym{mfcc}{MFCC}{Mel Frequency Cepstral Coefficient}
\newacronym{stft}{STFT}{Short Time Fourier Transform}
\newacronym{cqcc}{CQCC}{Constant Q Cepstral Coefficient}
\newacronym{roc}{ROC}{Receiver Operating Characteristic}
\newacronym{auc}{AUC}{Area Under the Curve}
\newacronym{tp}{TP}{True Positive}
\newacronym{tn}{TN}{True Negative}
\newacronym{fp}{FP}{False Positive}
\newacronym{fn}{FN}{False Negative}
\newacronym{tpr}{TPR}{True Positive Rate}
\newacronym{fpr}{FPR}{False Positive Rate}
\newacronym{tnr}{TNR}{True Negative Rate}
\newacronym{vae}{VAE}{Variational Autoencoders}
\newacronym{gan}{GAN}{Generative Adversarial Networks}
\newacronym{stlt}{STLT}{Short-Term and Long-Term}
\newacronym{gru}{GRU}{Gated recurrent unit}
\newacronym{ser}{SER}{Speech Emotion Recognition}
\newacronym{dft}{DFT}{Discrete Fourier Transform}
\newacronym{mae}{MAE}{Mean Absolute Error}
\newacronym{svm}{SVM}{Support Vector Machines}
\newacronym{gmm}{GMM}{Gaussian Mixture Models}
\newacronym{gnn}{GNN}{Graph Neural Network}
\newacronym{enf}{ENF}{Electric Network Frequency}
\newacronym{nlp}{NLP}{Natural Language Processing}
\newacronym{cm}{CM}{countermeasure}
\newacronym{mlp}{MLP}{Multi-Layer Perceptron}
\newacronym{ssl}{SSL}{Self-supervised Learning}
\newacronym{asv}{ASV}{Automatic Speaker Verification}
\newacronym{vad}{VAD}{Voice Activity Detection}
\newacronym{eer}{EER}{Equal Error Rate}

\begin{document}

\title{Forensic Similarity for Speech Deepfakes}

\author{Viola Negroni}
\email{viola.negroni@polimi.it}
\orcid{0009-0000-2483-4366}
\affiliation{%
  \institution{DEIB, Politecnico di Milano}
  \city{Milano}
  \country{Italy}
}

\author{Davide Salvi}
\email{davide.salvi@polimi.it}
\orcid{0000-0002-5163-3364}
\affiliation{%
  \institution{DEIB, Politecnico di Milano}
  \city{Milano}
  \country{Italy}
}

\author{Daniele Ugo Leonzio}
\email{danieleugo.leonzio@polimi.it}
\orcid{0000-0002-3217-9952}
\affiliation{%
  \institution{DEIB, Politecnico di Milano}
  \city{Milano}
  \country{Italy}
}

\author{Paolo Bestagini}
\email{paolo.bestagini@polimi.it}
\orcid{0000-0003-0406-0222}
\affiliation{%
  \institution{DEIB, Politecnico di Milano}
  \city{Milano}
  \country{Italy}
}

\author{Stefano Tubaro}
\email{stefano.tubaro@polimi.it}
\orcid{0000-0002-1990-9869}
\affiliation{%
  \institution{DEIB, Politecnico di Milano}
  \city{Milano}
  \country{Italy}
}

\renewcommand{\shortauthors}{Negroni et al.}

\begin{abstract}
In this paper, we introduce the concept of \textit{forensic similarity} in the speech deepfake detection domain, which aims to determine whether two audio segments share the same underlying forensic traces. 
Our approach is inspired by prior work in the image domain. 
To transfer this idea to the audio domain, we propose a two-stage deep learning framework consisting of a Siamese-based feature extractor and a core decision module, referred to as the similarity network. The system goal to assess whether two speech samples originate from the same source by comparing their forensic characteristics. In practice, the model maps pairs of audio segments to a similarity score indicating whether they contain identical or different forensic traces.
We evaluate the proposed method on the emerging task of source verification, demonstrating its ability to determine whether two speech samples were generated by the same model. In addition, we explore its applicability to audio splicing detection as a complementary use case. Experimental results show that the proposed approach generalizes well to previously unseen forensic traces, highlighting its robustness, flexibility, and practical relevance for digital audio forensics.
\end{abstract}



\begin{teaserfigure}
  \fbox{\includegraphics[width=\textwidth]{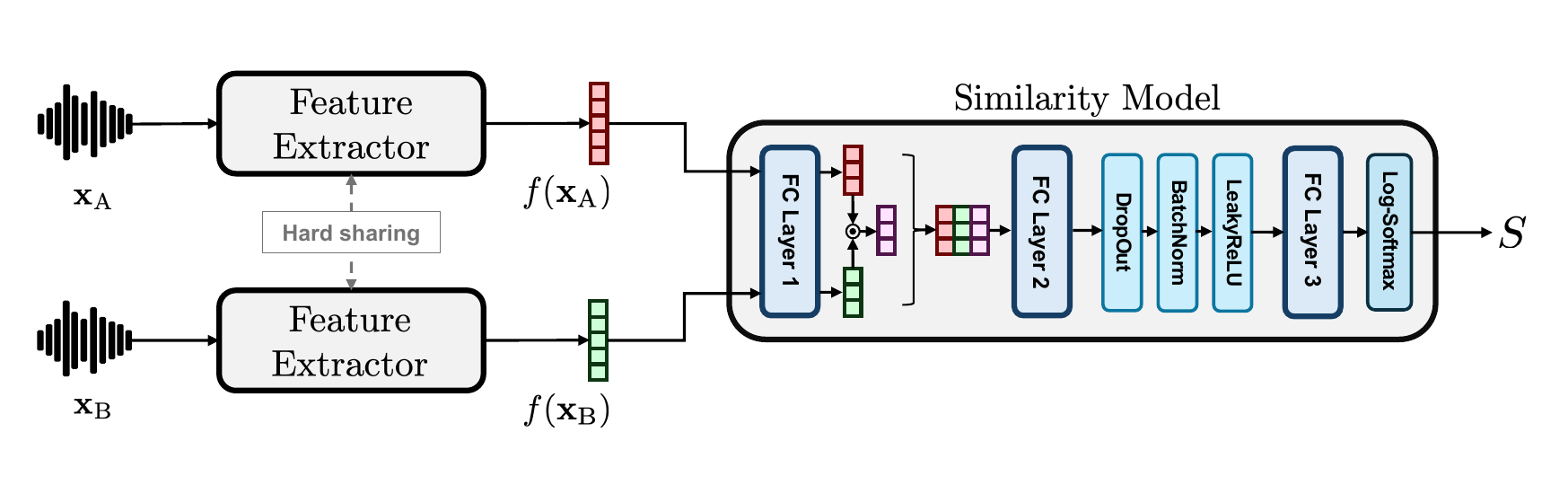}}
  \caption{Overview of the proposed framework.}
  \label{fig:teaser}
\end{teaserfigure}

\maketitle

\input{Sections/1_intro}
\input{Sections/2_method}
\input{Sections/3_setup}
\input{Sections/4_results}
\input{Sections/5_conclusion}

\begin{acks}

This work was supported by the FOSTERER project, funded by the Italian Ministry of Education, University, and Research within the PRIN 2022 program.
This work was partially supported by the European Union - Next Generation EU under the Italian National Recovery and Resilience Plan (NRRP), Mission 4, Component 2, Investment 1.3, CUP D43C22003080001, partnership on “Telecommunications of the Future” (PE00000001 - program “RESTART”).
This work was partially supported by the European Union - Next Generation EU under the Italian National Recovery and Resilience Plan (NRRP), Mission 4, Component 2, Investment 1.3, CUP D43C22003050001, partnership on ``SEcurity and RIghts in the CyberSpace’’ (PE00000014 - program ``FF4ALL-SERICS’’).

\end{acks}

\bibliographystyle{ACM-Reference-Format}
\bibliography{references}

\end{document}

%% file: Sections/1_intro.tex
\section{Introduction}
\label{sec:intro}

The increasing availability of AI-based generative technologies is significantly impacting multiple aspects of our daily lives.
In the information field, for instance, it is now possible to generate highly realistic synthetic media with minimal effort, blurring the line between authentic and synthetic content.
While these technologies open the doors to new exciting possibilities, they can also introduce serious threats when misused~\cite{amerini2025deepfake}.

A prominent example is the rise of \textit{deepfakes}, synthetic multimedia content generated through deep learning techniques that depict individuals in actions and behaviours that do not belong to them. Nowadays, anyone can fabricate images, videos, or audio of target victims, and these manipulated media have already been used for scams, fake news, damage reputations, and erode public trust.

In the audio domain, deepfakes can be exploited to impersonate a target speaker’s voice, making them say arbitrary utterances~\cite{ajder2019state, maras2019determining, vaccari2020deepfakes}.
To mitigate the risks related to the misuse of this content, the multimedia forensics community has been working to develop synthetic speech detection systems able to discriminate between real and fake audio content.
The task that such systems aim to solve is known as speech deepfake detection. 

From early signal processing techniques to modern deep learning architectures, speech deepfake detection has undergone a rapid evolution.
Initial progress was driven by convolutional models applied to mel-spectrograms, an approach adapted from image classification. Representative examples include ResNet~\cite{he2016deep} and LCNN~\cite{wu2018light}, which remain competitive and continue to reach state-of-the-art performance.
In parallel, other solutions emerged with SincNet~\cite{ravanelli2018speaker}, which introduced parametrized band-pass filters for learning task-specific filterbanks directly from raw waveforms. This principle laid the foundation for RawNet2~\cite{tak2021end}, where SincNet-inspired filterbanks were integrated with residual blocks and Gated Recurrent Units (GRUs) to form a robust end-to-end architecture. 
AASIST~\cite{jung2022aasist} advanced this approach with a heterogeneous graph attention framework capturing spectral and temporal artifacts.
More recently, research has increasingly shifted toward detectors built on top of large \gls{ssl} encoders~\cite{salvi2024comparative}, with Wav2vec 2.0~\cite{baevski2020wav2vec} and XLS-R~\cite{babu2021xlsr} being particularly influential in this transition.

Despite significant progress in speech deepfake detection, one of the primary challenges that the detectors need to face is generalization~\cite{muller2024harder, kawa2022attack}.
This refers to maintaining strong detection performance when evaluated against audio signals produced by generative models not encountered during training.
In fact, different generators leave behind distinctive artifacts, or forensic traces, that can be used to establish the authenticity of a given audio signal.
As these traces may differ substantially across systems, their variability can undermine the reliability of detectors trained on a limited set of sources.
This problem is also amplified by the rapid proliferation of increasingly sophisticated speech synthesis methods, which produce outputs that are both diverse and highly realistic~\cite{barakat2024deep}.
In response, recent research has shifted focus from simply determining whether a signal is authentic to analyzing its origin, aiming to attribute or verify the generative source behind the audio~\cite{bhagtani2023synthesized, mishra2026towards}.

While real–fake discrimination remains important, framing the task solely as a binary classification problem is increasingly restrictive. It collapses the diversity of generative models into a single ``fake’’ category, limiting robustness to unseen architectures and novel synthesis techniques. This not only weakens generalization in practical scenarios but also provides limited forensic value: knowing that a signal is synthetic is useful, but understanding \textit{how} it was produced or tracing it back to its source offers far greater insight for accountability and the design of effective countermeasures.

In this work, we focus on \textit{source verification}~\cite{negroni2025source}, a subtask of \textit{source tracing} (also known as synthetic speech attribution) that performs implicit attribution of the source of an audio through comparative verification.
Given two audio signals, the goal of source verification is to determine whether they were produced by the same generative model or by different ones.
Unlike closed-set attribution approaches, this strategy avoids the need to train explicitly on every possible generator, thereby improving scalability and robustness to new synthesis methods.

To this end, we introduce the concept of \textit{forensic similarity}~\cite{mayer2019forensic} in the speech deepfake detection domain and propose a novel framework that determines whether two audio segments share the same forensic trace (i.e., generator-specific artifacts).
Our method builds on prior work in image forensics, where forensic similarity has been shown to generalize to previously unseen traces without requiring explicit knowledge of them during training.

The proposed system follows a Siamese architecture, consisting of two main components:
(i) a feature extractor based on a source tracing backbone, which captures discriminative forensic cues, and
(ii) a lightweight similarity network that maps pairs of feature representations to a similarity score.
We evaluate the framework on source verification and benchmark it against different baselines, additionally showing its applicability to splicing detection as a real-world use case.
Results indicate that our approach generalizes effectively to unseen generative models, achieving reliable performance in both controlled and open-set conditions. These findings highlight the flexibility and practical value of forensic similarity as a tool for digital audio forensics.

The main contributions of this paper are the following:
\begin{itemize}
	\item We introduce the forensic similarity framework~\cite{mayer2019forensic} to the audio domain and adapt it for speech deepfake analysis.
	\item We deploy our system on the task of source verification, showing  improvements with respect to priors~\cite{negroni2025source}.
	\item We show how this system can be applied to the use case of splicing detection, suggesting a promising alternative approach to this task.
\end{itemize}

The rest of the paper is organized as follows.
\Cref{sec:background} reviews related work on source tracing for speech deepfakes. 
\Cref{sec:method} introduces the proposed framework, \Cref{sec:setup} describes the experimental protocol. 
\Cref{sec:results} reports and discusses the evaluation outcomes, while \Cref{sec:use_case} presents a real-world application to splicing detection in partially spoofed speech.
Finally, \Cref{sec:conclusions} summarizes our findings and outlines directions for future work.


\section{Related Work}
\label{sec:background}
In this section, we present an overview of the state of the art in source tracing for speech deepfake detection, where source verification is considered a sub-task. In addition, we review prior work on splicing detection in the speech deepfake domain (also referred to as partially fake speech detection) to provide the necessary background for understanding and framing the use case described in \Cref{sec:use_case}.

\subsection{Source Tracing and Source Verification}
\label{subsec:back_st}

In this section, we review current approaches to source tracing in speech deepfake detection, followed by an overview of source verification.

Synthetic speech attribution, or source tracing, aims to identify the generative model behind a given synthetic speech sample.
This task has become increasingly important for forensic investigations and content authenticity verification. However, progress in this area has been limited, with existing methods often struggling with generalization, scalability, and robustness in open-set scenarios, where test samples may come from previously unseen generators.

Early approaches to speech deepfake attribution treated the problem as a closed-set classification task, where models were trained to identify the source of a synthetic sample from a predefined set of known generators~\cite{borrelli2021synthetic, salvi2022exploring}. 
While effective under controlled conditions, these methods break down when tested against deepfakes produced by novel generators.
Closed-set models also face scalability challenges: as the number of deepfake generators grows, classifiers must be retrained frequently, a process that is impractical given the rapid pace of innovation in speech synthesis. 
These limitations make closed-set approaches unsuitable for real-world applications, where it is unrealistic to assume access to all possible generators.

An alternative approach analyzes the synthesis process itself rather than specific generators~\cite{klein2024source}. 
By extracting features from stages like acoustic models and vocoders, it can generalize across generators by detecting common patterns. 
However, as end-to-end deepfake systems become more complex, the reliability of this method for source attribution remains uncertain.

Recent work has increasingly focused on the challenges of open-set. 
TADA~\cite{stan2025tada} introduces a training-free approach based on k-Nearest Neighbors applied to self-supervised embeddings, achieving strong performance in both within-dataset attribution and out-of-domain detection. 
The authors of~\cite{koutsianos2025synthetic} evaluate classification- and metric learning-based approaches, showing that ResNet backbones can match or outperform \gls{ssl} models for generative system attribution while integrating speaker recognition techniques with audio forensics. 
The authors of~\cite{falez2025audio} propose open-set evaluation protocols, including few-shot identification and verification, while~\cite{klein2025open} introduces a softmax energy adaptation for out-of-distribution detection, demonstrating substantial improvements for multiple augmentation strategies. 

Most recently, the authors of~\cite{negroni2025source} propose a novel framework for synthetic speech analysis, reframing deepfake attribution as a source verification problem.
The method compares embeddings of a query track with those of reference signals to determine whether they originate from the same generative model. By operating in this verification paradigm, it avoids the need for exhaustive training on all possible generators and allows new reference sets to be incorporated flexibly, providing a scalable and generalizable approach to open-set source tracing.

\subsection{Splicing Detection}
\label{subsec:back_splicing}

In the audio domain, the area of splicing detection has remained relatively unexplored until recent times. 
This research field is closely related to audio copy-move, as they are the most generally used techniques that can alter the content of an audio track. 
Nevertheless, in audio copy-move, modifications are made by copy-pasting parts of a speech track within the very same speech track. 
In the case of splicing instead, the speech track is typically interleaved with audio patches that have a different origin with respect to the host track. 
In other words, a spliced signal consists of a speech track that has been manipulated by inserting short audio patches taken from elsewhere. 
These segments can stem from various origins, typically recordings of the same individual captured by different devices or synthetically generated speech segments, which are increasingly common.
The challenge in speech audio splicing detection is about determining whether a speech audio track originates from a single recording or if it is a fusion of two or more distinct tracks. 

In the field of synthetic speech detection, the focus has traditionally been on distinguishing completely genuine from completely spoofed audio tracks, and concerns about splicing attacks have arisen more recently. 
These efforts can be classified into two categories: transition boundary detection and segment-level classification. 
The Audio Deep Synthesis Detection challenge (ADD 2022)~\cite{yi2022add} is the first challenge attempting to tackle this kind of attack. 
In the context of this challenge, the authors of~\cite{wu2022partially} use the \gls{nlp} based technique of question-answering strategy with a self-attention mechanism to detect transition boundaries between pristine and fake segments. 
On the other hand,~\cite{cai2023waveform} proposes to use the self-supervised learning model wav2vec2.0~\cite{baevski2020wav2vec} for frame-level boundary detection. 
The authors of~\cite{wang2022synthetic} propose a SE-Res2Net-Conformer architecture to detect the spliced segment boundaries.
As for the segment-level approach, the goal here is to distinguish between genuine and fake segments at different time resolutions within the given track. 
Segments that only contain genuine speech will be labeled as 1, while all other segments will be labeled as 0. 
Zhang et al.~\cite{zhang2021initial} made a first attempt to perform segment-level classification for partially fake speech detection using a fixed time resolution and introduce a new database based on ASVspoof2019~\cite{todisco2019asvspoof}, named PartialSpoof, that was designed for this task. 
In their later works~\cite{zhang2021multi, zhang2022partialspoof}, they propose training the \gls{cm} model using both utterance-level and segment-level labels and employ self-supervised learning models as the front-end feature extractor. 
They also expand PartialSpoof by adding segment labels for various temporal resolutions. 
In~\cite{yadav2024mdrt}, the authors propose a system that uses transformer neural networks and processes multi-domain features using a ResNet-style \gls{mlp} to perform detection at multiple resolutions on PartialSpoof.

Nevertheless, both existing methods for segment-level classification and those aimed at detecting transition boundaries are meant for training on spliced tracks. 
We believe this might constitute a limitation for splicing detection methods intended for real-world applications, as analysts typically do not have access to huge databases of spliced tracks. 
To address this issue, we observe that a spliced track can be viewed as a patchwork of speech segments from different sources. 
Based on this observation, we apply our proposed framework to analyze consecutive portions of a test speech signal in a pairwise manner. 
The goal is to determine whether adjacent segments share the same origin (non-spliced) or come from different sources (spliced), and to evaluate whether this constitutes a promising alternative approach worth further exploration.

%% file: Sections/2_method.tex
\section{Proposed Method}
\label{sec:method}
This section formalizes the task of source verification and describes the proposed framework based on forensic similarity.

\subsection{Problem Formulation}
\label{subsec:system}

Let $\mathbf{x}_\text{A}$ and $\mathbf{x}_\text{B}$ be two synthetic speech samples. 
The goal of the source verification task is to determine whether the two audio signals originate from the same speech synthesis model.
The task can be formulated as a binary detection problem, where we need to predict the label $y^\text{SIM} \in \{0,1\}$ for the pair $(\mathbf{x}_\text{A}, \mathbf{x}_\text{B})$, as in
\begin{equation*}
y^\text{SIM} =
\begin{cases}
1, & \text{if } \mathbf{x}_\text{A} \text{ and } \mathbf{x}_\text{B} \text{ come from the same generator}, \\
0, & \text{otherwise}.
\end{cases}
\end{equation*}
Note that this formulation slightly differs from that in~\cite{negroni2025source}. 
There, an individual test track is compared against a set of \num{5} reference tracks generated by the same model. 
In contrast, our Siamese framework performs direct one-to-one comparisons between pairs of audio samples, which can be viewed as a special case of the reference-based approach where the reference set contains a single track.

\subsection{System Overview}
\label{subsec:system}
Our framework is inspired by the concept of forensic similarity introduced in~\cite{mayer2019forensic} and adapts it to the audio domain.
The fundamental idea of this approach is to assess whether two speech signals share the same forensic trace, which in our context corresponds to generation artifacts left by a particular speech synthesis model.
If the traces differ, the segments originate from different sources, and if they match, they share a common origin. 
Unlike traditional source tracing methods that aim to explicitly identify specific generators, forensic similarity is particularly suited to work in open-set scenarios, as it does not require prior exposure to a generation trace in order to make a prediction. 
Rather than identifying target forensic signatures, the system assesses whether the traces in two input signals are consistent. 
By focusing on relative similarity rather than absolute attribution, the approach remains effective even when the signals originate from previously unseen synthesis models.

The pipeline we propose consists of two modules:
\begin{enumerate}
 \item \textbf{Feature extractor}: A speech deepfake detection backbone that encodes each input into a forensic-meaningful embedding.
 \item \textbf{Similarity model}: A lightweight neural network that maps a pair of embeddings to a similarity score $S$.
\end{enumerate}
The two modules are trained sequentially: first, the feature extractor is optimized for source tracing, then its embeddings are used to train the similarity model in a Siamese setup~\cite{cozzolino2019noiseprint}. 

\Cref{fig:teaser} shows the complete architecture of the proposed framework: $\mathbf{x}_\text{A}$ and $\mathbf{x}_\text{B}$ are a pair of input speech samples, while $f(\mathbf{x}_\text{A})$ and $f(\mathbf{x}_\text{B})$ are, respectively, the embeddings we extract from these samples. 
An embedding is a dense representation of the input that captures the most informative and discriminative features learned by the model, i.e., the feature extractor, reflecting the patterns relevant to the task it was trained on.
The similarity model then computes the similarity score $S$, indicating how closely the two samples share the same forensic traces.
In the following, we present the two modules in detail.

\subsubsection{Feature Extractor}
\label{subsubsec:feature_extractor}
The feature extractor maps an input speech signal $\mathbf{x}$ into an embedding $f(\mathbf{x})$ capturing generator-specific cues. 
The framework is agnostic to the backbone, allowing different architectures to be plugged in. 
In this work, we experiment with four speech deepfake detection models: LCNN~\cite{wu2018light} and ResNet18~\cite{he2016deep}, both operating on mel-spectrograms, as well as RawNet2~\cite{tak2021end} and AASIST~\cite{jung2022aasist}. 
All of these models are well-established state-of-the-art anti-spoofing systems, that is, speech deepfake detectors. 
In this work, we repurpose them to perform closed-set source tracing, using them as embedding extractors to transform input audio into compact representations that capture the most relevant patterns for distinguishing between different generators.

Traditional source tracing, or deepfake attribution, is often framed as a multi-classification problem.
The objective is to identify the specific generator used to synthesize fake samples.
This approach encourages the network to learn deeper and more informative representations of the input, while a diverse training set further enhances its ability to generalize to unseen generators in open-set scenarios.
To train the considered networks for source tracing, we set the size of their final fully connected layer to match the number of classes, i.e., synthetic speech generators, in the training dataset, following the approach of~\cite{salvi2022exploring}. 
Then, embeddings are extracted from the last hidden layer of the feature extractor, and the trained feature extractor is used to construct a Siamese configuration that feeds these embeddings into the similarity model. 
This setup employs hard sharing, meaning both branches use the same weights and biases.
Since the feature extractor is applied twice, the first stage produces a pair of embeddings for two input speech samples, $f(\mathbf{x}_{A})$ and $f(\mathbf{x}_{B})$, which are then passed to the similarity model.

\subsubsection{Similarity Model}

The similarity model is a shallow neural network that predicts how similar two embeddings are in terms of forensic traces.
Its inputs are the feature vectors $f(\mathbf{x}_\text{A})$ and $f(\mathbf{x}_\text{B})$ of dimension $L$, extracted in the previous step from the speech segments $\mathbf{x}_A$ and $\mathbf{x}_B$.
The network performs source verification by producing a similarity score $S \in [0,1]$, where higher values indicate stronger correspondence between the traces. 
In this source verification setting, a higher score implies a greater likelihood that both segments were generated by the same model.

Each input embedding is first projected to a lower-dimensional space through a fully connected layer, producing vectors $h_\text{A}$ and $h_\text{B}$ of size $M < L$:
\begin{equation}
h_\text{A} = W_\text{fc1} f(\mathbf{x}_\text{A}) + b_\text{fc1}, \quad
h_\text{B} = W_\text{fc1} f(\mathbf{x}_\text{B}) + b_\text{fc1}.
\end{equation}
Note that the architecture of the first layer of the similarity model is adapted to the size $L$ of the input vector, depending on the chosen feature extractor.

To enhance the network’s expressive capacity, we perform a concatenation operation following the strategy in~\cite{mayer2019forensic}. 
Specifically, we concatenate the two outputs of the fully connected layer along with a third vector obtained via element-wise multiplication of the two, as in 
\begin{equation}
h_\text{prod} = h_\text{A} \odot h_\text{B} \in \mathbb{R}^M,
\end{equation}
\begin{equation}
h_\text{concat} = [h_\text{A} \, \| \, h_\text{B} \, \| \, h_\text{prod}] \in \mathbb{R}^{3M}.
\end{equation}
The resulting concatenated vector is then processed through an additional fully connected layer that brings $h_\text{concat}$ back to $M$:
\begin{equation}
h_\text{out} = W_\text{fc2} h_\text{concat} + b_\text{fc2} \in \mathbb{R}^M.
\end{equation}

Then, we incorporate a dropout layer, batch normalization, and a leaky ReLU activation function to improve the generalization and performance. 
The final layer is a fully connected layer followed by a LogSoftmax that computes log probabilities for two classes, indicating whether the two audio patches exhibit similar forensic traces or not:
\begin{equation}
S = \text{LogSoftmax}(W_\text{fc3} h_\text{out} + b_\text{fc3}) \in \mathbb{R}^2.
\end{equation}

Finally, the similarity score $S$ is converted into a hard decision by applying a threshold $\tau$, producing the prediction $\hat{y}^\text{SIM}$, which is an estimate of the similarity label $y^\text{SIM}$:
\begin{equation}
\hat{y}^\text{SIM} = 
\begin{cases} 
1 & \text{if } S \ge \tau \\
0 & \text{if } S < \tau. 
\end{cases}
\end{equation}
A prediction of $\hat{y}^\text{SIM}=1$ indicates that the two segments are considered to share the same forensic trace, implying a common generative source. Conversely, $\hat{y}^\text{SIM}=0$ denotes that the traces differ, suggesting that the segments were produced by different generators.

%% file: Sections/3_setup.tex
\section{Experimental Setup}
\label{sec:setup}
In this section we present the datasets employed within this study. 
We then detail the experimental setup we used, together with the technical choices that led to the results presented in~\Cref{sec:results} and~\Cref{sec:use_case}.

\subsection{Evaluation Datasets}
\label{subsec:datasets}
We now describe the datasets employed within this study. 
For all the considered data we assumed a sampling rate equal to \SI{16}{\kilo\hertz}.

\noindent\textbf{MLAAD \cite{muller2024mlaad}}.
This is a large-scale dataset that includes only synthetic speech signals, generated using \num{82} \gls{tts} models across \num{33} distinct architectures.
It was designed to evaluate anti-spoofing systems in multi-language and multi-generator scenarios.
We use the fifth version of this corpus, which contains \num{378} hours of synthetic speech in \num{38} different languages, and follow the source tracing protocols (i.e., data splits) proposed in~\cite{UsingMLAADforSourceTracing}. 
This dataset serves as the primary training corpus for our experiments. Unless otherwise specified, model training is conducted using a combination of the \textit{training} and \textit{development} partitions, which encompass \num{41} synthetic speech generators. 
Evaluation is performed on the \textit{test set} and is restricted to generators that do not overlap with the training and development splits, enabling open-set analysis, again totaling \num{41} generators.

\noindent\textbf{ASVspoof 2019 \cite{todisco2019asvspoof}}.
This dataset was developed for the homonymous challenge, aiming to push research towards the development of more effective \gls{asv} systems. 
We employ the Logical Access partition of this corpus.
It includes real speech from the VCTK corpus~\cite{veaux2016superseded} and synthetic speech fabricated by \num{17} different synthetic speech generators. The data are distributed unevenly across partitions to support open-set evaluation: the \textit{train} and \textit{dev} sets share spoofed utterances generated with \num{6} algorithms, while the \textit{eval} set is disjoint and contains spoofed utterances generated with \num{11} unseen algorithms. 

\noindent\textbf{TIMIT-TTS \cite{salvi2023timit}}.
This is a synthetic speech dataset derived from the VidTIMIT corpus~\cite{sanderson2002vidtimit}. For our experiments, we use its \textit{clean} partition, which contains speech signals generated using \num{12} different \gls{tts} models, with \num{7} dedicated to single-speaker synthesis and \num{5} supporting multiple speakers.
The single-speaker partition replicates the voice of Linda Johnson from LJSpeech~\cite{ljspeech17}, while the multi-speaker partition synthesizes voices from LibriSpeech~\cite{panayotov2015librispeech}. 

\noindent\textbf{PartialSpoof~\cite{zhang2022partialspoof}}
This is an English speech database derived from the ASVspoof 2019 LA corpus~\cite{todisco2019asvspoof}, which contains both real and partially fake speech signals.
It follows the same structure as the ASVspoof dataset and is divided into three partitions: training, development, and evaluation. 
This design enables models to be trained directly on spliced tracks, offering an alternative approach for tackling the splicing detection task compared to the one we are proposing here.
The dataset includes synthetic speech generated by \num{17} distinct methods, varying between training-development and evaluation subsets.
PartialSpoof has been designed to detect even the shortest audio splices, which is why it includes fake segments that can be as short as a few milliseconds. 

To summarize, we train our framework on MLAAD and evaluate it in-domain using the disjoint subset of its test partition to measure generalization to unseen generation algorithms (\textit{in-domain, open-set scenario}). 
To further examine \textit{out-of-domain} generalization, we test on ASVspoof 2019 and TIMIT-TTS. 
Please note that, unlike MLAAD and TIMIT-TTS, ASVspoof 2019 includes genuine speech. 
Since real samples are not present in MLAAD and thus never observed during training, we treat them as an additional class only at evaluation time.

As for the use case of splicing detection, we train on ASVspoof 2019 training set rather than PartialSpoof, as our method does not require training on spliced data. 
This also serves to expose the model to genuine speech that is absent in MLAAD.
Splicing-specific performance is then assessed on the PartialSpoof development and evaluation sets, with the development set used to demonstrate \textit{closed-set} evaluation of seen generation algorithms, and the evaluation set employed to test generalization in an \textit{open-set} scenario.

\subsection{First Learning Phase - Feature Extractor}
\label{subsec:fe_setup}
We trained the feature extractors for closed-set source tracing in a supervised manner. 
For each model, this required adjusting the number of outputs in the final fully connected layer, originally set to \num{2} for the detection task, to match the number of classes in our training dataset, corresponding to \num{41} generation algorithms. 
The networks were fed speech segments along with their associated class labels $y^\text{GEN}_c$, $c \in [0, 1, \dots, C-1]$, $C$ is the total number of classes in the training set.

We use input speech segments of duration \SI{4}{\second}, which is common practice in the literature. 
Shorter segments are padded by repeating the signal, while longer utterances are truncated to fit the required length. During training, each audio patch is sampled with equal probability from any of the generators in the training set, following standard practice to mitigate class imbalance. 
We merged the training and development sets of MLAAD and randomly split them into \SI{70}{\percent} for training and \SI{30}{\percent} for validation, while preserving the class distribution across generation algorithms.

For LCNN, RawNet2, and ResNet18, we trained each network for \num{200} epochs using a batch size of \num{256}. 
We employed the Cross Entropy loss function with the Adam optimizer and an initial learning rate of \num{0.001}, reduced on plateau based on the validation loss. The scheduler patience was set to \num{10} epochs, and early stopping to \num{20} epochs. 
These measures help prevent the models from getting stuck in local minima and reduce the risk of overfitting.
AASIST was trained with a batch size of \num{64}. 
All other training settings, including scheduler and early stopping, were the same as above.
For ResNet18 and LCNN, mel-spectrograms were computed using a Fourier transform size of \num{512}, a window length of \num{400}, and a hop length of \num{160}. 
The frequency range was restricted to $f_\text{min} = 20\ \text{Hz}$ and $f_\text{max} = 7600\ \text{Hz}$, a Hamming window function was applied, and the number of mel bins was set to \num{80}.

\subsection{Second Learning Phase - Similarity Model}
\label{subsec:sm_setup}
In this second training stage, we focus on training the similarity model to learn a forensic similarity mapping, i.e., to determine whether two speech samples share the same origin or come from different sources. 
The pre-trained feature extractor is imported and arranged in the Siamese configuration described in \Cref{sec:method}, while the training data remain the same as in \Cref{sec:setup}~B. 

When using RawNet2 as the feature extractor, the embedding dimensionality is $L = 1024$. 
For ResNet18, the embeddings have size $L = 256$, for AASIST $L = 160$, and for LCNN $L = 128$. 
In all cases, the dimensionality is reduced to $M = 64$ through the initial fully connected layer of the similarity model, which was selected based on experimental tuning.

We conducted parallel experiments under two training strategies: in the first, the weights of the feature extractor are kept frozen; in the second, the extractor is fine-tuned jointly with the similarity model whenever the latter improves. The similarity model receives pairs of embeddings generated by the Siamese architecture from two selected input speech tracks, together with a binary label (\num{1} if both embeddings belong to the same class $y^\text{GEN}_c$, \num{0} otherwise). 

The network was trained for \num{100} epochs with a batch size of \num{256} (\num{64} when the feature extractor is AASIST), using Negative Log Likelihood loss, Adam optimizer, and an initial learning rate of \num{0.0001}. 
The scheduler patience and early stopping criteria were the same as in the previous training stage.

\subsection{Source Verification Protocol}
\label{subsec:sv_setup} 
All test tracks are adjusted to a fixed duration of \SI{4}{\second}, as for training. 
Following standard practice, the central window is selected for testing on longer signals.
An exhaustive evaluation is then performed by comparing each test sample against every other sample in the test set.
As similarity scores are computed using a neural network, they are not guaranteed to be symmetric. 
Consequently, for each pair of samples $\mathbf{x}_\text{A}$ and $\mathbf{x}_\text{B}$, the model produces two scores corresponding to the ordered inputs $(\mathbf{x}_\text{A}, \mathbf{x}_\text{B})$ and $(\mathbf{x}_\text{B}, \mathbf{x}_\text{A})$. 
These scores are averaged to obtain a single, order-invariant similarity value for the pair.

\subsection{Splicing Detection Setup}
\label{subsec:splicing_setup}
For this scenario, the system is provided with pairs of sequential signal windows extracted from a single input audio signal at a time.
The number of window pairs analyzed varies with the length of each speech track, as consecutive analysis windows are slid across the track until its end is reached.

From each test track, speech pairs of \SI{0.5}{\second} each are extracted using two consecutive, non-overlapping sliding windows with a stride of \SI{0.05}{\second}, generating overlapping pairs.
Remaining trailing fragments that do not fit the windowing scheme are discarded.
Inference is performed with a batch size of \num{1}, allowing the generation of a similarity score sequence for each test utterance.

\begin{figure}
    \centering
    \includegraphics[width=1\columnwidth]{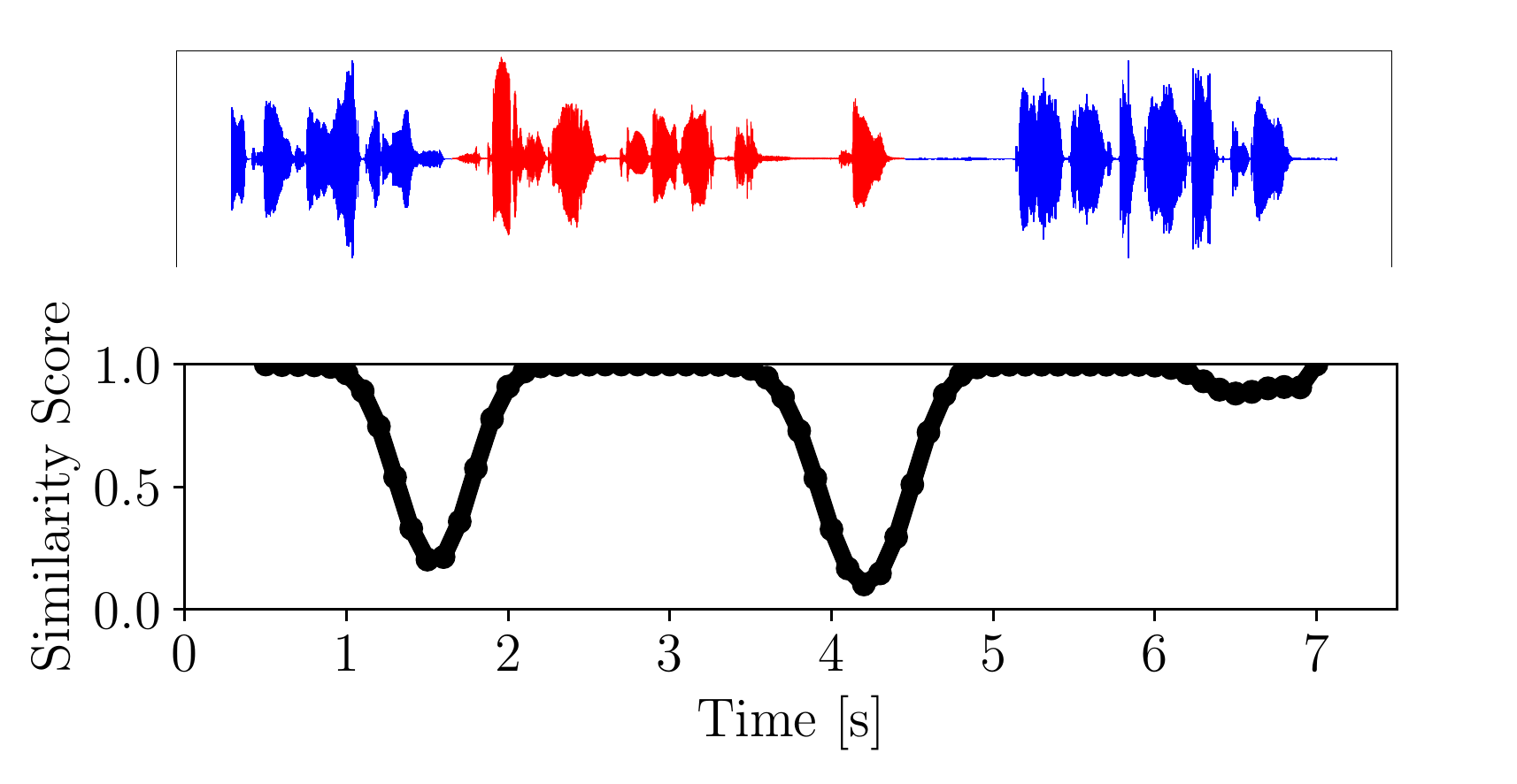}
    \caption{Example of similarity score sequence computed by the proposed framework for the splicing detection use case over a partially spoofed track (blue segments indicate genuine speech, red synthetic).}
    \label{fig:splicing_example}
\end{figure}

Specifically, adjacent window pairs are sequentially fed into the Siamese framework, which computes similarity scores $S$ between them.
For each pair, embeddings are extracted via the feature extractor and passed through the similarity model, producing eventually a raw similarity score sequence along the track.
To reduce noise, this sequence is smoothed using a Gaussian filter with $\sigma = 1.7$.
\Cref{fig:splicing_example} shows an example of this process.

Similarity scores $S$ range from \num{1} (high similarity) to \num{0} (low similarity), with lower scores indicating greater dissimilarity between the forensic traces of the analyzed segments.
Therefore, potential splice boundaries are identified as local minima in the smoothed sequence, with minima defined by a minimum depth of \num{0.38} and a minimum width of \num{3} signal samples.
These parameters were tuned exclusively on the PartialSpoof training set, which was never used for model training nor evaluation (we used a \num{80}/\num{20} train/validation split on ASVspoof 2019 train set, see \Cref{sec:setup}~A).
Finally, the largest (in absolute value) detected minimum is taken as the global score for the track and used to determine whether the track is authentic or partially spoofed, i.e., contains spliced segments.

%% file: Sections/4_results.tex
\section{Results}
\label{sec:results}
In this section, we present the results of our experimental campaign.
First, we conduct an exhaustive experiment to identify the best-performing feature extractor and determine which is the most effective training strategy for source verification.
Next, we benchmark the optimal configuration against alternative similarity assessment approaches, testing on previously unseen generators in both in-domain and out-of-domain scenarios.
Finally, we provide a detailed analysis of the in-domain source verification results, examining performance on a per-generator basis to gain deeper insights.

\subsection{Front-End Feature Extractor and Training Strategy}
\label{subsec:exp_1}

\begin{table}
\centering 
\caption{Comparison of feature extractors under different training strategies. Open-set results in terms of EER and AUC on the MLAAD test set.}
\label{table:fe_ts}
\begin{tabular}{cccc}
\toprule 
\textbf{Feature Extractor} & \textbf{Training Strategy} & \textbf{EER $\downarrow$} & \textbf{AUC $\uparrow$} \\ \midrule \midrule
\multirow{2}{*}{RawNet2} & Frozen   & 23.8\% & 83.7\% \\
                          & Unfrozen & 24.3\% & 83.2\% \\ \midrule
\multirow{2}{*}{ResNet18} & Frozen   & 12.1\% & 94.0\% \\
                           & Unfrozen & \textbf{10.4\%} & 95.3\% \\ \midrule
\multirow{2}{*}{\textbf{LCNN}}     & Frozen   & 11.6\% & 94.6\% \\
                           & \textbf{Unfrozen} & 10.5\% & \textbf{95.7\%} \\ \midrule
\multirow{2}{*}{AASIST}   & Frozen   & 21.8\% & 86.7\% \\
                           & Unfrozen & 19.6\% & 88.0\% \\ \bottomrule
\end{tabular}
\end{table}

This experiment is designed to identify the most effective feature extractor and assess whether it performs better when kept frozen or fine-tuned during the second learning phase.
\Cref{table:fe_ts} reports the source verification performance of the tested configurations in terms of \gls{eer} and \gls{auc} on \num{41} unseen, in-domain generators from MLAAD.
Among the tested feature extractors, LCNN and ResNet18 perform best at distinguishing whether test samples originate from the same generator. 
While both perform similarly when fine-tuned, LCNN slightly outperforms ResNet18 in AUC when unfrozen. Based on these results, we select the LCNN with an unfrozen training strategy as the front-end feature extractor for the subsequent experiments.

\subsection{Similarity Function Benchmarking}

\begin{table*}
\caption{Contribution to the framework of the similarity model on the source verification task compared against other similarity scoring methods on in-domain (MLAAD open-set) and out-of-domain (TIMIT-TTS and ASVspoof2019) test sets. Results are reported in terms of EER and AUC.}
\label{tab:sim_bench}
\centering
\begin{tabular}{ccccccc|cc}
\hline
\toprule
\multirow{2}{*}{} & \multicolumn{2}{c}{\textbf{MLAAD}} & \multicolumn{2}{c}{\textbf{TIMIT-TTS}} & \multicolumn{2}{c}{\textbf{ASVspoof 2019}} & \multicolumn{2}{c}{\textbf{Average}} \\ \cmidrule(lr){2-3} \cmidrule(lr){4-5} \cmidrule(lr){6-7} \cmidrule(lr){8-9}
                  & \textbf{EER $\downarrow$} & \textbf{AUC $\uparrow$}  & \textbf{EER $\downarrow$} & \textbf{AUC $\uparrow$}  & \textbf{EER $\downarrow$} & \textbf{AUC $\uparrow$} & \textbf{EER $\downarrow$} & \textbf{AUC $\uparrow$}  \\ \midrule
Euclidean Distance     & 19.9  & 87.7  & 34.1  & 73.0  & 27.9  & 79.5  & 27.3  & 80.1  \\
Cosine Distance~\cite{negroni2025source} & 18.6  & 88.9  & 33.0  & 74.0  & 28.6  & 78.7  & 26.7  & 80.5  \\
Contrastive Learning~\cite{chopra2005learning}  & 18.4  & 89.6  & 32.8  & 75.5  & 27.5  & \textbf{80.4}  & 26.2  & 81.8  \\ 
\textit{Similarity Model (ours)}   & \textbf{10.5}  & \textbf{95.7}  & \textbf{31.1}  & \textbf{77.3}  & \textbf{25.6}  & 78.8  & \textbf{22.4}  & \textbf{83.9}  \\ \bottomrule
\end{tabular}
\end{table*}

In this experiment, we assess the validity of the similarity model for source verification by comparing it with standard similarity scoring methods.
To do so, we systematically replace it within the framework with cosine similarity~\cite{negroni2025source}, Euclidean distance, and a contrastive learning approach commonly used in Siamese networks~\cite{chopra2005learning}.
Note that, unlike our similarity model and contrastive learning, cosine and Euclidean distances do not require a second learning stage.
We report performance on unseen generators from MLAAD (in-domain, open-set) as well as out-of-domain unseen generators from TIMIT-TTS and ASVspoof2019, the latter including real speech as a standalone class.
As shown in \Cref{tab:sim_bench}, the similarity model consistently outperforms the other approaches across all datasets and evaluation metrics, achieving the lowest EERs and highest AUCs on average. 
This highlights its strength in capturing subtle source-specific details while maintaining robust generalization.

\subsection{Source-Level Verification Analysis}

\begin{figure*}
    \centering
    \includegraphics[width=\textwidth]{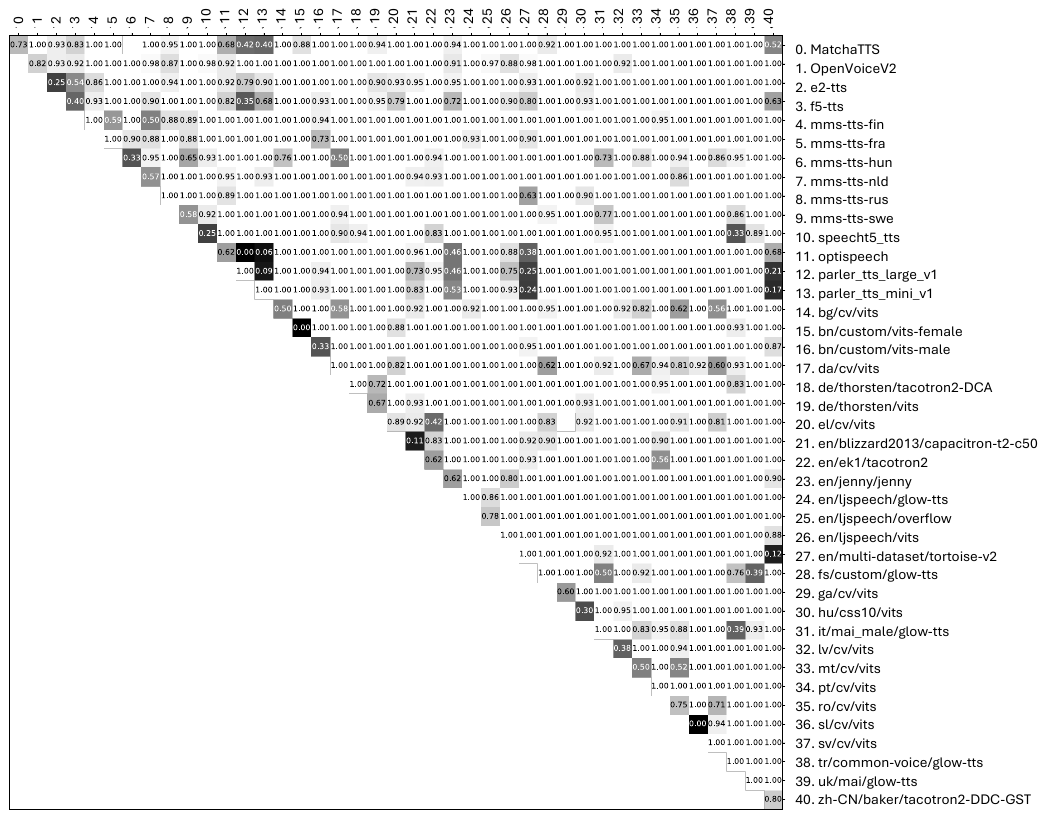}
    \caption{Detection rates of the proposed framework using LCNN as the feature extractor, with fine-tuning applied during the second learning phase, evaluated for each generator pair on the MLAAD test set. Lighter cells indicate higher detection rates.}
    \label{fig:matrix}
\end{figure*}

We now assess the source verification performance of our proposed framework on MLAAD on a per-generator basis.
\Cref{fig:matrix} shows the detection rates for each generator pair, using a threshold $\tau$ that was tuned on the validation set.
The diagonal entries show the correct classification rates when two speech samples come from the same generator.
The non-diagonal entries of the matrix show the correct classification rates of when the samples come from different generators.

Overall, the framework achieves near-perfect accuracy for the majority of generator pairs, successfully identifying whether speech samples originate from the same model, even when the underlying synthesis methods were not included during training.
However, certain generators prove more challenging than others. 
For instance, Parler TTS is occasionally confused with a few other generators. 
Likewise, speech from some non-English generators is misclassified as originating from different sources, even when produced by the same model, as seen with MMS TTS (Hungarian and Dutch) and several VITS models trained on languages such as Latvian, Slovenian, and Bulgarian.
These errors may arise from subtle intra-linguistic differences or from the scarcity of training data available for less-represented languages, which can introduce variability in the generated speech that may interfere with or hinder the generation traces, making it harder for the framework to recognize them. 


\section{Use Case: Application to Splicing Detection}
\label{sec:use_case}
In this section, we evaluate the potential of the proposed framework for the practical task of splicing detection. 

The problem of splicing detection is attracting growing attention due to the increasing availability of editing tools and advances in speech-generation technologies. 
Splicing can be performed by cutting and pasting words from different speeches of the same individual, but modern deepfake generators simplify this process, eliminating the need for a large corpus of target speech.

Here, we focus on detecting audio tracks where parts of a real utterance are replaced with synthetically generated segments.
First, we test whether our framework can maintain performance on shorter input signals to determine its robustness for splicing detection under small-window conditions. 
Next, we assess its splicing detection performance on the PartialSpoof dataset.

\subsection{Impact of Signal Duration}

\begin{table}
\caption{Source verification results vs. input duration from models trained on ASVspoof 2019 train set, evaluated on dev + eval sets (EER and AUC).}
\label{tab:input_duration}
\centering
\begin{tabular}{ccc}
\toprule
\textbf{Win Length (s)} & \textbf{EER $\downarrow$} & \textbf{AUC $\uparrow$} \\ \midrule
4.0   & 14.5 & 92.8 \\
3.0   & 16.7 & 91.2 \\
2.0   & 18.3 & 90.4 \\
1.0   & 22.3 & 86.6 \\
0.5 & 24.9 & 84.1 \\ \bottomrule
\end{tabular}
\end{table}

This preliminary experiment aims to assess how system performance evolves as the duration of input signals decreases.
For this investigation, the best-performing framework from \Cref{sec:results}~A is retrained on ASVspoof 2019 to ensure exposure to genuine speech. 
For each considered window length, only the second training phase, i.e., similarity model training, is repeated with the shortened window.
As expected, performance gradually degrades as the input window shortens (\Cref{tab:input_duration}). Nevertheless, even at an input length of \SI{0.5}{\second}, the framework still achieves an AUC of $84.1\%$, which demonstrates a notable degree of robustness to limited input and highlights its potential applicability to the splicing detection task.

\begin{figure}
    \centering
    \includegraphics[width=0.8\columnwidth]{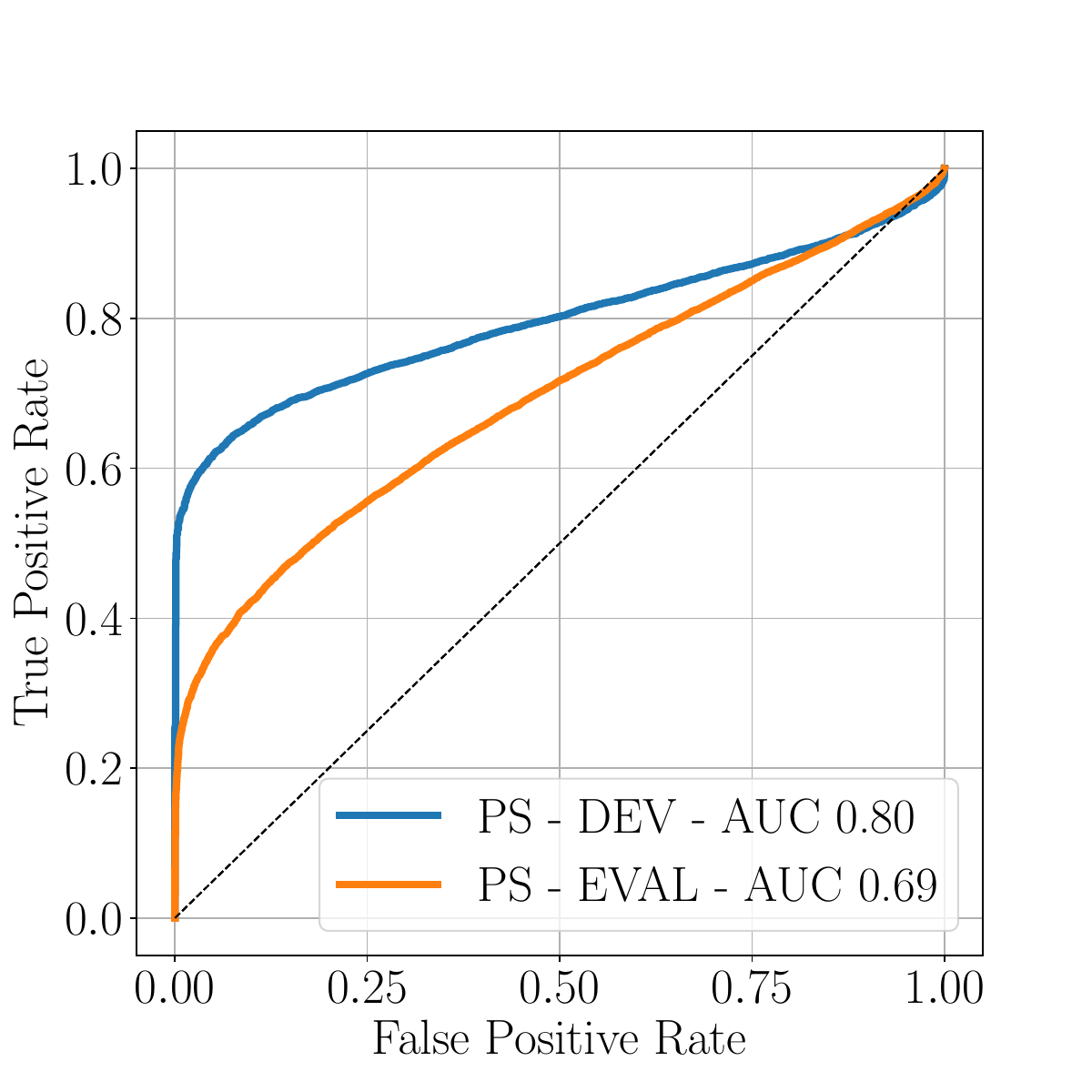}
    \caption{ROC curves for splicing detection on PartialSpoof using input pairs of 0.5 seconds with a stride of 0.05 seconds.}
    \label{fig:roc_ps}
\end{figure}

\subsection{Splicing Detection Performamce}
The results of this experiment are reported in \Cref{fig:roc_ps}. 
On the PartialSpoof development set, which uses the same generators seen during training, the framework achieves an \gls{auc} of $80\%$. 
On the PartialSpoof evaluation set, which includes \num{11} unseen generators (compared to \num{6} seen during training plus real samples), the \gls{auc} drops to $69\%$.
This outcome highlights clear room for improvement and likely reflects limited generalization due to the small number of generators used in training.

Notably, both ROC curves initially rise steeply at low false positive rates.
This means that the model correctly identifies many spliced signals while rarely misclassifying genuine signals. 
Such a conservative behavior with respect to false positives is indeed desirable from a forensic perspective, as it minimizes the risk of erroneously flagging authentic content.

%% file: Sections/5_conclusion.tex
\section{Conclusions and Future Work}
\label{sec:conclusions}

In this work, we introduced forensic similarity to the audio domain, proposing a Siamese-based framework for source verification of speech deepfakes. 
By focusing on the comparison of forensic traces rather than relying solely on binary real–fake discrimination, our approach demonstrated strong generalization to previously unseen generative models and proved effective in both source verification and splicing detection tasks. 
The results highlight the value of treating deepfake forensics as a similarity-driven problem, offering a scalable and robust approach that consistently improves over existing baselines for open-set source verification.

As future work, we aim to refine the framework to better mitigate the influence of linguistic and speaker-specific cues. 
We also plan to further tailor it for splicing detection, where preliminary results are promising, and to explore its potential for precise splicing point localization.
We also plan to explore the integration of more powerful detection models equipped with self-supervised learning (SSL) front-ends to enhance feature extraction.
This might be particular beneficial for further exploring the potential of the method for the splicing detection task, which in particular may further boost performance on splicing detection tasks.